\documentstyle[twocolumn,prl,aps,epsfig]{revtex}
\begin{document}
\flushbottom \draft
\title{Atomic four-wave mixing: fermions versus bosons}
\author{M. G. Moore and P. Meystre}
\address{Optical Sciences Center\\
University of Arizona, Tucson, Arizona 85721\\ (April 20, 1999)
\\ \medskip}\author{\small\parbox{14.2cm}{\small \hspace*{3mm}
We compare the efficiency of four-wave mixing in quantum
degenerate gases of bosonic and fermionic atoms. It is shown that
matter-wave gratings formed from either bosonic or fermionic atoms
can in principle exhibit nearly identical Bragg-scattering, i.e.
four-wave mixing, properties. This implies that effects such as
coherent matter-wave amplification and superradiance can occur in
degenerate fermi gases. While in the boson case the Bragg
resonance is clearly due to `Bose enhancement', in the case of
fermions the resonance is due to constructive many-particle
quantum interference.
\\[3pt]PACS numbers: 03.75.Fi, 03.75-B}}
\maketitle \narrowtext

The experimental realization of {\em nonlinear atom optics}
\cite{LenMeyWri93} is one of many recent advances made possible by
the achievement of atomic Bose-Einstein condensation. The
demonstration of atomic four-wave mixing \cite{DenHagWen99}, the
discovery of BEC superradiance \cite{InoChiSta99,MooMey99b}, and
the development of coherent matter-wave amplification
\cite{InoPfaGup99,KozSuzTor99} are all examples of nonlinear wave
mixing involving atomic Bose-Einstein condensates. One proposed
interpretation of these phenomena, in which the generation of a
new atomic and/or optical side-modes \cite{MooMey99a} is
attributed to `Bose-enhancement', predicts that these effects
would not be seen in Fermi systems. In contrast, a second
interpretation which attributes the effect to Bragg scattering
from atomic matter-wave gratings, predicts that four-wave mixing
with fermions should, in fact, be possible.

In this letter we address this fundamental issue and compare the
Bragg-scattering properties of quantum-degenerate bosonic and
fermionic matter-wave gratings. Three issues need to be addressed:
First, there is the question of whether it is indeed possible  to
create a high-quality matter-wave grating in a quantum degenerate
Fermi gas. Second, one needs to determine the scattering
properties from such a grating. Lastly, a precise
quantum-mechanical interpretation of atomic four-wave mixing must
be formulated. Our analysis shows that the establishment of a
fermionic grating is indeed possible, and furthermore leads to
scattering properties practically indistinguishable from those of
a BEC with the same mean density profile. We then present a
physical interpretation of this result, contrasting the {\em Bose
enhancement} responsible for four-wave mixing in the boson case to
a many-particle {\em quantum interference} effect in the fermion
case.

We consider a system consisting of two scalar fields with
annihilation operators denoted as $\hat{\mit\Psi}_1({\bf r})$ and
$\hat{\mit\Psi}_2({\bf r})$. The first field contains $N$
identical bosonic or fermionic atoms from which a matter-wave
grating will be formed. The second field contains a single test
particle which will probe the scattering properties of the
grating. This test particle might be an atom, in which case the
scattering properties are related to atomic four-wave mixing
experiments, or it could be a photon, in which case the results
would be applicable to phenomena such as BEC superradiance and
matter-wave amplification. The commutation relations for the two
fields are therefore given by
\begin{eqnarray}
    \hat{\mit\Psi}_i({\bf r})\hat{\mit\Psi}^\dag_j({\bf r}')
    -\mu_{ij}\hat{\mit\Psi}^\dag_j({\bf r}')\hat{\mit\Psi}_i({\bf r})
    &=&\delta_{ij}\delta^3({\bf r}-{\bf r}')\nonumber\\
    \hat{\mit\Psi}_i({\bf r})\hat{\mit\Psi}_j({\bf r}')
    -\mu_{ij}\hat{\mit\Psi}_j({\bf r}')\hat{\mit\Psi}_i({\bf r})
    &=&0,
\label{comm1}
\end{eqnarray}
where the coefficients $\mu_{ij}$ vary depending on the quantum
statistics of the two fields. If both fields are bosonic we have
$\mu_{ij}=1$. If, however, the grating is fermionic then either
the test particle is an atom of the same species in a different
internal state, in which case $\mu_{ij}=-1$, or it is a different
type of bosonic (fermionic) particle, in which case $\mu_{11}=-1$,
$\mu_{12}=\mu_{21}=1$, and $\mu_{22}=1(-1)$.

The two fields are subject to the free Hamiltonians $\hat{\cal
H}_1$ and $\hat{\cal H}_2$, respectively, and are coupled via a
two-body interaction of the form
\begin{equation}
    \hat{\cal V}=\lambda \int d^3r\, \hat{\mit\Psi}^\dag_1({\bf r})
    \hat{\mit\Psi}^\dag_2({\bf r})
    \hat{\mit\Psi}_2({\bf r})\hat{\mit\Psi}_1({\bf r}),
\label{V}
\end{equation}
which describes equally well atom-atom collisions in the $s$-wave
scattering approximation or the effective interaction between
ground-state atoms and far off-resonant photons. For simplicity we
neglect collisions between atoms in the grating.

At present, the primary technique for establishing density
gratings in BEC is by use of a 'beam splitter' based on two-photon
Bragg transitions between center-of-mass states
\cite{KozDenHag99}. Each atom is thereby transferred into a
coherent superposition of its initial state and a copy of that
state displaced in momentum space by the two-photon recoil kick.
Quantum interference between these two momentum groups then
results in a typical `standing wave' density modulation.

For the purpose of this letter we specialize to the case of
gratings formed from harmonically confined gases at zero
temperature. We first introduce the 3-dimensional harmonic
oscillator states
\begin{equation}
    \varphi_m({\bf r})=\phi_{\alpha(m)}(x/\xi_x)
    \phi_{\beta(m)}(y/\xi_y)\phi_{\gamma(m)}(z/\xi_z),
\label{phim3d}
\end{equation}
where $\xi_j$ is the oscillator length along the $j$-axis,
$\alpha(m)$, $\beta(m)$, and $\gamma(m)$ are the quantum numbers
of the $m$th 3-d oscillator energy level, and $\phi_n$ is the
normalized $n$th 1-d harmonic oscillator energy level. We further
define the creation operators $\hat{a}^\dag_m({\bf k})$ for the
{\em momentum side-modes} of the $m$th oscillator state as
\begin{equation}
    \hat{a}^\dag_m({\bf k})=\int d^3r\, \varphi_m({\bf r})e^{i{\bf k}\cdot{\bf r}}
    \hat{\mit\Psi}^\dag_1({\bf r}).
\label{amk}
\end{equation}
At $T=0$ the states of $N$-atom Bose and Fermi gases are
$[\hat{a}^\dag_0(0)]^N|0\rangle/\sqrt{N!}$ and
$\prod^{N-1}_{m=0}\hat{a}^\dag_m(0)|0\rangle$, respectively. Thus
immediately after the Bragg pulse is applied, the state of the
matter-wave grating is given for bosons by
\begin{equation}
    |\psi_B\rangle=[2^NN!]^{-1/2}
    \left[\hat{a}^\dag_0(0)+e^{i\theta}\hat{a}^\dag_0({\bf K})\right]^N|0\rangle
\label{psiB}
\end{equation}
and for fermions by
\begin{equation}
    |\psi_F\rangle=[2^N]^{-1/2}\prod^{N-1}_{m=0}
    \left[\hat{a}^\dag_m(0)+e^{i\theta}\hat{a}^\dag_m({\bf K})\right]|0\rangle,
\label{psiF}
\end{equation}
where $\theta$ and $\hbar{\bf K}$ are the relative phase and
momentum transfer imparted on the atoms by the Bragg splitter.
Henceforth we assume $\theta=0$ for simplicity. We note that these
states are only normalized in the limit that $\hbar{\bf K}$ is
much larger than the rms spread in momentum of the initial Bose
and Fermi clouds, a condition which we assume to hold. It is this
condition, in fact, which is sufficient to guarantee a
high-quality grating at wavelength $2\pi/K$.

To compare the Bragg-induced gratings in degenerate Bose and Fermi
gases we compare mean atomic densities $\rho({\bf
r})=\langle\hat{\mit\Psi}^\dag_1({\bf r})\hat{\mit\Psi}_1({\bf
r})\rangle$ for the two states (\ref{psiB}) and (\ref{psiF}). By
making use of the commutation relations (\ref{comm1}) we find
\begin{equation}
    \rho_B({\bf r})=N|\varphi_0({\bf r})|^2[1+\cos({\bf K}\cdot{\bf r})],
\label{rhoB}
\end{equation}
and
\begin{equation}
    \rho_F({\bf r})=\sum^{N-1}_{m=0}|\varphi_m({\bf r})|^2[1+\cos({\bf K}\cdot{\bf r})].
\label{rhoF}
\end{equation}
which shows that in both cases, the effect of the Bragg splitter
is to superpose a density modulation $[1+\cos({\bf K}\cdot{\bf
r})]$ to the mean density of the initial atomic cloud. Fig. 1
shows the integrated density $\rho(x,z)=\int dy\, \rho({\bf r})$
for the case of spherically symmetric harmonic traps with
oscillator lengths $\xi$ for the Fermi gas and
$\xi_b=\sqrt{2}\Delta x_f$ for the Bose gas, where $\Delta x_f$ is
the rms spread of the initial Fermi cloud. This choice of $\xi_b$
guarantees that the rms widths of the two clouds are equal in
${\bf r}$-space. The parameters for Fig.1 are ${\bf
K}=(20/\xi)\hat{z}$ and $N=86$, and $x$ and $z$ are given in units
of $\xi$. Figure 1a shows the result for the BEC, while Fig. 1b
shows the result for the degenerate Fermi gas.

This example establishes that it is possible to construct very
similar matter-wave gratings from degenerate Bose and Fermi gases.
It should be noted that while their mean densities are nearly
identical in coordinate space, they are very different in momentum
space. As the phase-space density of the BEC is much higher than
that of the degenerate Fermi gas, the bosonic grating is
significantly more localized in momentum space.\\ \vspace{1 ex}

The next step is to compare the scattering properties of these two
gratings. Of particular interest here is the issue of Bose
enhancement, which is oftentimes invoked as the cause of
stimulated scattering in bosons. In order to address this
question, we consider the case where a single test particle is
incident on the matter-wave with some wave-vector ${\bf k}_0$ and
use perturbation theory to compute the probability $P({\bf k},t)$
that it is scattered into a given state ${\bf k}$ after some time
$t$. The initial state of the system is therefore taken as
\begin{equation}
    |\psi(0)\rangle=\hat{c}^\dag({\bf k}_0)|\psi_{B,F}\rangle,
\label{psi0}
\end{equation}
where
\begin{equation}
    \hat{c}^\dag({\bf k}_0)=V^{-1/2}\int d^3r\, e^{i{\bf k}_0\cdot{\bf r}}
    \hat{\psi}^\dag_2({\bf r})
\label{ck}
\end{equation}
creates a plane-wave test particle of momentum ${\bf k}_0$ in the
quantization volume $V$. The probability at time $t$ of finding
the test particle in the plane-wave state ${\bf k}$ is therefore
\begin{equation}
    P({\bf k},t)=\langle\psi(t)|\hat{c}^\dag({\bf k})\hat{\cal P}_N
    \hat{c}({\bf k})|\psi(t)\rangle,
\label{Pkt}
\end{equation}
where $|\psi(t)\rangle$ is the solution of the Schr\"odinger
equation
\begin{equation}
    |\psi(t)\rangle=\exp\left[-i(\hat{\cal H}_1+\hat{\cal H}_2+\hat{\cal V})t/\hbar\right]
    |\psi(0)\rangle
\label{psit}
\end{equation}
and $\hat{\cal P}_N$ is the $N$-atom projection operator
\begin{eqnarray}
    \hat{\cal P}_N&=&\int d^3r_1\ldots d^3r_N\,
    \hat{\mit\Psi}^\dag_1({\bf r}_N)\ldots
    \hat{\mit\Psi}^\dag_1({\bf r}_1)|0\rangle\nonumber\\
    &\times&\langle0|
    \hat{\mit\Psi}_1({\bf r}_1)\ldots\hat{\mit\Psi}_1({\bf r}_N).
\end{eqnarray}

This scattering problem can be solved analytically by expanding
the solution (\ref{psit}) to first order in the parameter
$\lambda$ of Eq. (\ref{V}). This yields the scattering
probabilities after the test particle as been scattered once by
the grating. In order to proceed we first specify the free
Hamiltonians $\hat{\cal H}_1$ and $\hat{\cal H}_2$. Our approach
is to assume that the states $\varphi_m({\bf r})\exp[i{\bf
k}\cdot{\bf r}]$ are approximate eigenstates of the
first-quantized version of $\hat{\cal H}_1$, with energy $E_m({\bf
k})\approx=\hbar\omega_m+\hbar^2k^2/2M$ where $\hbar\omega_m$ is
the energy of the $m$th trap eigenstate and $M$ is the atomic
mass. These states are essentially low lying levels of the the
harmonic trap shifted in momentum space by $\hbar{\bf k}$. For
ultracold atoms it is reasonable to neglect the evolution of the
wavepackets for times small compared to the oscillator length
divided by the velocity $\hbar k/m$. For these short times the
overlap between the initial wavepacket and the state it evolves
into remains approximately unity. In addition, we take the
plane-waves $(1/\sqrt{V})\exp[i{\bf k}\cdot{\bf r}]$ to be
eigenstates of the first-quantized version of $\hat{\cal H}_2$
with energy $\hbar^2k^2/2m$.

An analytic expression for $P({\bf k},t)$ can be derived in a
relatively straightforward manner.  For ${\bf k}\neq{\bf k}_0$ and
$|{\bf K}-{\bf k}_0|\sim K$ we find that the leading order (in
$\lambda$) contribution to Eq. (\ref{Pkt}) is given for the boson
case by
\begin{eqnarray}
    P_B({\bf k},t)&\approx&
    \frac{\lambda^2N}{2\hbar^2V^2}\left[
    |F_1({\bf k},t)|^2+|F_2({\bf k},t)|^2\right. \nonumber\\
    &+&(N-1)|F_1({\bf k},t)+F_2({\bf k},t)|^2
    |G_{00}({\bf k}-{\bf k}_0)|^2\nonumber\\
    &+&\left. (N-1)|F_2({\bf k},t)|^2|G_{00}({\bf k}-{\bf k}_0-{\bf K})|^2\right],
\label{PBkt}
\end{eqnarray}
and for the fermion case by
\begin{eqnarray}
    P_F({\bf k},t)&\approx&\frac{\lambda^2N}{2\hbar^2V^2}\left[
    |F_1({\bf k},t)|^2+|F_2({\bf k},t)|^2\right. \nonumber\\
    &+&|F_1({\bf k},t)+F_2({\bf k},t)|^2{\cal G}({\bf k}-{\bf k}_0)\nonumber\\
    &+&\left. |F_2({\bf k},t)|^2{\cal G}({\bf k}-{\bf k}_0-{\bf K})\right],
\label{PFkt}
\end{eqnarray}
where the time-dependent functions
\begin{equation}
    F_1({\bf k},t)=\frac{\sin\left[\frac{\hbar}{2m}({\bf k}-{\bf k}_0)\cdot{\bf k}t\right]}
    {\left[\frac{\hbar}{2m}({\bf k}-{\bf k}_0)\cdot{\bf k}\right]}
    e^{i\frac{\hbar}{2m}({\bf k}-{\bf k}_0)\cdot{\bf k}t}
\label{F1}
\end{equation}
and
\begin{equation}
    F_2({\bf k},t)=\frac{\sin\left[\frac{\hbar}{2m}({\bf k}-{\bf k}_0)\cdot({\bf k}-{\bf K})t\right]}
    {\left[\frac{\hbar}{2m}({\bf k}-{\bf k}_0)\cdot({\bf k}-{\bf K})\right]}
    e^{i\frac{\hbar}{2m}({\bf k}-{\bf k}_0)\cdot({\bf k}-{\bf K})t}
\label{F2}
\end{equation}
give the effects of energy conservation, and the functions
\begin{equation}
    G_{mn}({\bf k})=\frac{1}{\sqrt{2}}\int d^3r\, \varphi^\ast_m({\bf r})e^{i{\bf k}\cdot{\bf r})}
    \varphi_n({\bf r})
\label{Gmn}
\end{equation}
and
\begin{equation}
    {\cal G}({\bf k})=\frac{1}{N}\left[\left|\sum_{m=0}^{N-1}G_{mm}({\bf k})\right|^2
    -\sum_{m,n=0}^{N-1}\left|G_{mn}({\bf k})\right|^2\right]
\end{equation}
describe the shapes of the Bragg resonances. The approximation
indicates that we have dropped several (negligible) terms from the
exact expression.\\ \vspace{1 ex}

The first two terms in Eqs. (\ref{PBkt}) and (\ref{PFkt})
correspond to spontaneous scattering, whereas the third and fourth
terms correspond to $0$th order (small angle) and $1$st order
Bragg resonances. In Figure 2 we plot the scattering probabilities
in the $k_x$-$k_z$ plane for the case ${\bf k}_0=K\hat{x}$. Figure
2a shows the relative probabilities for the boson matter-wave
grating of Fig. 1a and Fig. 2b shows those of the fermion grating
of Fig. 1b. We have chosen $t=.025 (2m\xi^2/\hbar)$ so that the
Bragg resonance is slightly sharper than the energy conservation
envelope. The large spot at $k_z=0$ corresponds to small-angle
($0$th order) scattering, while the smaller spot at $k_z=20$ to
the $1$st order Bragg resonance.  The inset figures show close-ups
of the $1$st order resonances, i.e. the four-wave mixing signal.
We see, perhaps surprisingly, that these two microscopically very
different gratings produce almost identical scattering
cross-sections. This strongly suggests that (at least for large N)
it is the mean atomic density alone which determines the
efficiency of four-wave mixing processes. We note that for $N=1$
there is no Bragg-resonance, only spontaneous scattering. This
shows that the Bragg resonance is a {\em many particle} effect,
simply having a modulated mean density is not sufficient in the
case of a single-atom grating.

While both types of matter-wave gratings result in practically
identical scattering distributions, the underlying physical
processes which produce the Bragg resonances seem at first glance
to be quite different. We can gain considerable physical intuition
about them by considering the simple case $N=2$ in more detail. In
this case, the states of the Bose and Fermi gratings are
\begin{eqnarray}
    |\psi_B\rangle&=&\frac{1}{2\sqrt{2}}
    \left[\sqrt{2}\frac{1}{\sqrt{2}}\left[\hat{a}^\dag_0(0)\right]^2|0\rangle
    +2\hat{a}^\dag_0(0)\hat{a}^\dag_0({\bf K})|0\rangle\right.\nonumber\\
    &+&\left.\sqrt{2}\frac{1}{\sqrt{2}}\left[\hat{a}^\dag_0({\bf K})\right]^2|0\rangle\right],
\label{psiB2}
\end{eqnarray}
and
\begin{eqnarray}
    |\psi_F\rangle&=&\frac{1}{2}\left[\hat{a}^\dag_0(0)\hat{a}^\dag_1(0)|0\rangle
    +\hat{a}^\dag_0(0)\hat{a}^\dag_1({\bf K})|0\rangle\right. \nonumber\\
    &+&\left.\hat{a}^\dag_0({\bf K})\hat{a}^\dag_1(0)|0\rangle
    +\hat{a}^\dag_0({\bf K})\hat{a}^\dag_1({\bf K})|0\rangle\right],
\label{psiF2}
\end{eqnarray}
where we have written each term as an amplitude times a normalized
state. In this representation the gratings can be thought of as
quantum superpositions of different initial states. If the test
particle is scattered into state ${\bf k}$, then one of the atoms
in the grating must absorb a recoil momentum kick of $\hbar({\bf
k}_0-{\bf k})$. This is accomplished by acting on the states
(\ref{psiB2}) and (\ref{psiF2}) with the operator $\int d^3r\,
\hat{\mit\Psi}^\dag_1({\bf r})\exp[i({\bf k}_0-{\bf k})\cdot{\bf
r}]\hat{\mit\Psi}_1({\bf r})$. When the Bragg condition ${\bf
k}={\bf k}_0+{\bf K}$ is satisfied, the resulting final states of
the gratings are
\begin{eqnarray}
    |\psi_B\rangle'&=&\frac{1}{\sqrt{2}}\left[
    \hat{a}^\dag_0(-{\bf K})\hat{a}^\dag_0(0)|0\rangle
    +\hat{a}^\dag_0(-{\bf K})\hat{a}^\dag_0({\bf K})|0\rangle\right. \nonumber\\
    &+&\left. \sqrt{2}\frac{1}{\sqrt{2}}\left[\hat{a}^\dag_0(0)\right]^2|0\rangle
    +\hat{a}^\dag_0(0)\hat{a}^\dag_0({\bf K})|0\rangle\right],
\label{psiB2'}
\end{eqnarray}
where the factor $\sqrt{2}$ on the third term can be identified
with `Bose enhancement', and
\begin{eqnarray}
    |\psi_F\rangle'&=&\frac{1}{2}\left[
    \hat{a}^\dag_0(-{\bf K})\hat{a}^\dag_1(0)|0\rangle
    +\hat{a}^\dag_0(0)\hat{a}^\dag_1(-{\bf K})|0\rangle\right. \nonumber\\
    &+&\hat{a}^\dag_0(-{\bf K})\hat{a}^\dag_1({\bf K})|0\rangle
    +\hat{a}^\dag_0(0)\hat{a}^\dag_1(0)|0\rangle\nonumber\\
    &+&\hat{a}^\dag_0(0)\hat{a}^\dag_1(0)|0\rangle
    +\hat{a}^\dag_0({\bf K})\hat{a}^\dag_1(-{\bf K})|0\rangle\nonumber\\
    &+&\left. \hat{a}^\dag_0(0)\hat{a}^\dag_1({\bf K})|0\rangle
    +\hat{a}^\dag_0({\bf K})\hat{a}^\dag_1(0)|0\rangle\right].
\label{psiF2'}
\end{eqnarray}
In either case, the total transition probability is the sum of the
squares of the amplitudes for all distinguishable ``paths''. If
two final states are indistinguishable (e.g., the fourth and fifth
terms of (\ref{psiF2'})) then the amplitudes must be added before
taking the square modulus and adding to the other paths.

In both cases, therefore, the peak relative probability is $5/2$,
but in one case, it is attributed to Bose enhancement and in the
other to constructive quantum interference. We observe that if the
Bragg condition strongly violated, the Bose-enhancement factor
disappears from $|\psi_B\rangle'$, and that term is replaced by
two distinguishable states with weights equal to the other states.
In the fermionic case, on the other hand, the two
indistinguishable states become distinguishable. In both cases,
however, the resulting relative probability is reduced to $2$.

This example illustrates that the Bragg resonance in the case of a
BEC can be attributed to Bose enhancement, whereas in the
fermionic case it can be attributed to quantum interference
between ``paths'' which lead to indistinguishable final states.
That both mechanisms lead to precisely the same enhancement, a
fact which is readily verified for $N=3$ and illustrated for
$N=86$ by Fig. 2, is due to the fact that Bose enhancement, when
viewed in first-quantized many-particle quantum mechanics, is
simply a constructive quantum interference where many initial
states (the different terms under exchange of particle labels)
lead to a single final state (when all atoms are in same state
there is only one term even after symmetrization). Hence, effects
which can be interpreted as Bragg-scattering from atomic-matter
wave gratings, such as atomic four-wave mixing, BEC superradiance,
and matter-wave amplification, can work as efficiently {\it in
principle} in both Bose and Fermi systems. Because of the
connection between Bose enhancement and quantum interference we
find that, remarkably, it is possible to set up $N$-atom
superposition states in Fermi systems which mimic the effects of
Bose stimulation.

This work is supported in part by the U.S.\ Office of Naval
Research under Contract No.\ 14-91-J1205, by the National Science
Foundation under Grant No.\ PHY98-01099, by the U.S.\ Army
Research Office, and by the Joint Services Optics Program.

\begin{figure}
\caption{The integrated mean density $\rho(x,z)$ is shown for the
case of degenerate Bose (Fig. 1a) and Fermi (Fig. 1b) matter-wave
gratings. The parameters are $N=86$ and ${\bf K}=(20/\xi)$, and
the dimensionless variables $x$ and $\xi$ have been scaled to the
oscillator length $\xi$. }
\end{figure}
\begin{figure}
\caption{The relative scattering probabilities are shown for the
case ${\bf k}_0=K\hat{x}$ and $k_y=0$. Figures 2a and 2b
correspond to boson and fermion matter-wave gratings,
respectively. The inset figures show close-ups of the $1$st order
Bragg-resonances. The dimensionless parameters $k_x$ and $k_z$
have been scaled to $1/\xi$.}
\end{figure}

\end{document}